# HVPE Homoepitaxy of β-$Ga_2O_3$: High-Temperature Growth and Orientation-Dependent Surface Evolution

Sourav Sarker, Saleh Ahmed Khan, Ahmed Ibreljic, Pronoy Kanti Das, Anhar Bhuiyan [a)]

Department of Electrical and Computer Engineering, University of Massachusetts Lowell, Lowell, MA 01854, USA

[a)] Corresponding author Email: anhar_bhuiyan@uml.edu

**Abstract**

In this work, we systematically investigate homoepitaxial β-$Ga_2O_3$ growth by halide vapor phase epitaxy (HVPE) on (010), (001), and ($\bar{2}01$) native substrates over a broad growth parameter space, extending the growth temperature as high as 1150 °C. The growth kinetics exhibit strong dependence on precursor supply, temperature, reactor pressure, and Ga-source-to-substrate distance. Increasing the HCl flow from 80 to 300 sccm enhances the growth rate by up to seven-fold, while increasing the temperature from 1050 to 1150 °C nearly doubles the deposition rate. At 1150 °C, atmospheric-pressure growth yields rates more than 2.5 times higher than those at 100 Torr, while reducing the $O_2$ flow to 5 sccm further increases the growth rate to 14.3 μm/h, producing films up to 42.9 μm thick in 3 h. Notably, phase-pure homoepitaxial layers with good crystalline quality are obtained across all three orientations, with low rocking-curve full width at half maximum (FWHM) values of 63.4, 66.6, and 42.8 arcsec for the (010), (001), and ($\bar{2}01$) epilayers, respectively. The surface morphology exhibits a pronounced crystallographic dependence and evolves strongly with the HVPE growth conditions. In particular, the ($\bar{2}01$) surface develops well-defined step-and-terrace morphology with local RMS roughness as low as 0.17 nm. These results extend β-$Ga_2O_3$ HVPE homoepitaxy into a higher-temperature regime and demonstrate a broad process window for high-rate growth of thick, high-quality epilayers with strongly orientation-dependent surface evolution.

***Keywords:*** *Ultra-wide bandgap semiconductor, Halide Vapor Phase Epitaxy (HVPE),* β-$Ga_2O_3$*, homoepitaxy, surface morphology, crystallographic orientation*

β-$Ga_2O_3$ has emerged as a promising ultrawide bandgap semiconductor for next-generation high-power devices due to its large bandgap (4.8-4.9 eV), high critical electric field (6-8 MV/cm), and the availability of large-area melt-grown native substrates [1]. These attributes have enabled rapid advances in high-voltage β-$Ga_2O_3$ device demonstrations [2-17], with performance continuing to improve toward the material limits. The development of vertical device structures requires thick, low-defect epitaxial drift layers with high crystalline quality and controlled doping. A variety of epitaxial techniques have already been employed for β-$Ga_2O_3$ growth, including metal-organic chemical vapor deposition (MOCVD) [18-40], molecular beam epitaxy (MBE) [41-45], low-pressure chemical vapor deposition (LPCVD) [46-58], mist chemical vapor deposition (mist-CVD) [59-61], and halide vapor phase epitaxy (HVPE) [62-66]. Among these techniques, HVPE is particularly attractive for thick β-$Ga_2O_3$ epitaxy because of its high growth rate and capability for producing high-purity, low-doped layers. Early thermodynamic studies identified GaCl and $O_2$ as suitable precursors for β-$Ga_2O_3$ HVPE epitaxy [67], followed by the demonstration of high-purity homoepitaxial growth on (001) native substrates at rates up to 5 μm/h [68]. HVPE has since demonstrated thick (001) epilayers with high crystalline quality, including layers exceeding 50 μm in thickness with low n-type doping ($<1 \times 10^{16}$ cm$^{-3}$) on 2-inch and 100-mm substrates [65, 69].

The monoclinic structure of β-$Ga_2O_3$ results in strong crystallographic anisotropy, making the growth surface an important consideration in epitaxy. This sensitivity to the growth surface is also evident in heteroepitaxy. In our recent HVPE study on c-plane sapphire, intentional substrate offcut produced pronounced changes in epitaxial evolution and improved the structural and electrical properties, with the highest electron mobility obtained under the optimized offcut condition [70]. For homoepitaxy, while much of the early HVPE development focused on (001) substrates, growth has subsequently been demonstrated on several other crystallographic orientations. Leach et al. demonstrated homoepitaxy on (010) substrates and showed that the surface morphology is strongly influenced by substrate miscut [71]. Oshima et al. demonstrated growth on ($\bar{1}02$) substrates at rates up to 23 μm/h [72] and subsequently extended growth to the (011) orientation [64]. More recently, Butenko et al. reported orientation-dependent HVPE homoepitaxy on (100), ($\bar{2}01$), and (010) substrates [73]. While these studies demonstrate the influence of crystallographic orientation, the response of different β-$Ga_2O_3$ orientations to systematic changes in the HVPE growth environment remains largely unexplored. Growth temperature, precursor supply, and reactor pressure can strongly influence deposition kinetics and precursor transport. Gas-phase reactions between Ga-containing chloride species and oxygen can further influence precursor delivery and epitaxial growth. Notably, most β-$Ga_2O_3$ HVPE homoepitaxy has been investigated at temperatures near or below 1050 °C, leaving the higher-temperature growth regime unexplored. Here, we extend β-$Ga_2O_3$ HVPE homoepitaxy into this regime, reaching growth temperatures up to 1150 °C, while systematically comparing (010), (001), and ($\bar{2}01$) native substrates over a broad range of growth conditions. By varying HCl flow, Ga-source-to-substrate distance, growth temperature, $O_2$ flow, and reactor pressure, we examine the growth kinetics and orientation-dependent behavior of the three crystallographic surfaces, establishing a systematic understanding of their response across an extended HVPE growth window.

Homoepitaxial β-$Ga_2O_3$ films were grown on (010), (001), and ($\bar{2}01$) oriented β-$Ga_2O_3$ substrates. Prior to growth, the substrates were sequentially cleaned in acetone, isopropyl alcohol (IPA), and deionized (DI) water, followed by nitrogen drying. The HVPE growth process utilized high-purity metallic gallium (99.99999%) as the gallium source. HCl was introduced at flow rates ranging from 80 to 400 sccm and reacted with metallic Ga to generate volatile gallium chloride species for transport to the growth surface. $O_2$ was used as the oxidizing precursor at flow rates ranging from 5 to 20 sccm. The growth temperature was varied from 1050 to 1150 °C, while the reactor pressure was varied from 100 Torr to 760 Torr. The Ga-source-to-substrate distance was varied from 7.7 to 11.7 cm to investigate its influence on the growth kinetics. The crystalline structure, phase purity, orientation, and crystalline quality of the epilayers were evaluated using high-resolution X-ray diffraction (XRD) performed on a Rigaku SmartLab diffractometer equipped with Cu Kα radiation (λ = 1.5418 Å). The surface morphology of the as-grown epilayers was characterized using field-emission scanning electron microscopy (FESEM, JEOL JSM-7401F). Atomic force microscopy (AFM, Park XE-100) was used to examine the nanoscale surface morphology and determine the root-mean-square (RMS) surface roughness.

Figure 1 shows the influence of HCl flow rate and growth temperature on the growth kinetics of $\beta$-$Ga_2O_3$ films on (010), (001), and ($\bar{2}01$) $\beta$-$Ga_2O_3$ substrates. As shown in Fig. 1(a), the growth rate exhibits a strong dependence on HCl flow and Ga-source-to-substrate distance. At a growth temperature of 1050 °C and atmospheric pressure, an HCl flow of 80 sccm resulted in growth rates of 1.0-1.5 μm/h across the three substrate orientations for Ga-source-to-substrate distances of 7.7-8.6 cm, producing films with thicknesses of 3.0-4.5 μm after 3 h of growth. Increasing the HCl flow to 300 sccm substantially increased the growth rates to 7.4-9.3 μm/h at comparable source-to-substrate distances of 7.9-8.7 cm, corresponding to film thicknesses of 22.2-27.9 μm. Similar increases were observed for the (010), (001), and ($\bar{2}01$) substrates, with the growth rate increasing by up to seven-fold relative to that obtained at 80 sccm HCl. Such an increase in growth rate with HCl flow is likely associated with an increased supply of volatile Ga-containing chloride species generated through the reaction of HCl with the metallic Ga source. The Ga-source-to-substrate distance also has a pronounced influence on the growth rate. At an HCl flow of 400 sccm, growth rates of 4.6-4.8 μm/h were obtained for the (010), (001), and ($\bar{2}01$) substrates when positioned 11.1-11.7 cm from the Ga source, corresponding to film thicknesses of 13.8-14.4 μm after 3 h of growth. These rates are substantially lower than the 7.4-9.3 μm/h obtained at shorter distances of 7.9-8.7 cm using 300 sccm HCl, despite the higher HCl flow. This behavior indicates that increasing the HCl supply alone does not necessarily result in a higher deposition rate and that precursor transport from the Ga source to the substrate strongly affects the growth kinetics. The longer transport distance likely reduces the effective concentration of Ga-containing growth species reaching the substrate surfaces.

The influence of growth temperature is shown in Fig. 1(b) for films grown simultaneously on the three substrate orientations at 760 torr using HCl and $O_2$ flow rates of 400 and 20 sccm, respectively. At 1050 °C, growth rates of 4.6-4.8 μm/h were obtained across the (010), (001), and ($\bar{2}01$) substrates, corresponding to film thicknesses of 13.8-14.4 μm. Increasing the temperature to 1100 °C resulted in a modest increase, with growth rates reaching ~5.1 μm/h and film thicknesses of ~15.3 μm. A substantial increase was observed at 1150 °C, where the growth rates reached 8.7-9.1 μm/h, producing films with thicknesses of 26.1-27.3 μm. Increasing the growth temperature from 1050 to 1150 °C therefore nearly doubled the deposition rate. The enhanced growth rate at the higher temperature is likely associated with increased surface reaction and incorporation kinetics of the transported Ga-containing species.

The effects of $O_2$ flow and growth pressure on the deposition rates were further investigated at 1150 °C, as shown in Fig. 2. At an HCl flow of 400 sccm and atmospheric pressure, decreasing the $O_2$ flow from 20 to 5 sccm resulted in a substantial increase in growth rate [Fig. 2(a)]. At 20 sccm $O_2$, growth rates of 8.7-9.1 μm/h were obtained across the (010), (001), and ($\bar{2}01$) substrates, yielding film thicknesses of 26.1-27.3 μm after 3 h of growth. When the $O_2$ flow was reduced to 5 sccm, the growth rates increased to 11.7-14.3 μm/h, producing substantially thicker films of 35.1-42.9 μm. The highest growth rate of 14.3 μm/h, corresponding to a film thickness of 42.9 μm, was achieved under the lowest investigated $O_2$ flow. This inverse dependence of growth rate on $O_2$ flow suggests that excess oxygen suppresses the net deposition rate under the investigated HVPE conditions. The decrease in growth rate with increasing $O_2$ flow could be associated with enhanced gas-phase reactions between the Ga-containing chloride species and oxygen, which reduce the effective flux of growth species reaching the substrate surface. A strong dependence on growth pressure is also observed, as shown in Fig. 2(b). At 760 Torr, growth rates of 8.7-9.1 μm/h were obtained across the three substrate orientations, corresponding to film thicknesses of 26.1-27.3 μm. Reducing the pressure to 100 Torr decreased the growth rates to only 3.2-3.3 μm/h and the corresponding film thicknesses to 9.7-9.8 μm. The growth rate at atmospheric pressure is therefore more than 2.5 times higher than that at 100 Torr. A similar pressure dependence has been reported by Leach et al. for HVPE-grown $\beta$-$Ga_2O_3$, where decreasing the growth pressure from 100 to 65 Torr reduced the growth rate from 4.7 to 3.1 μm/h. The pressure dependence was associated with changes in gas velocity and vapor-phase reactions within the growth zone [74].

The structural properties of the homoepitaxial $\beta$-$Ga_2O_3$ layers were examined by high-resolution XRD, as shown in Fig. 3. The ω-2θ scans in Fig. 3(a) exhibit only the expected $\beta$-$Ga_2O_3$ reflections associated with the respective substrate orientations. For the (010) epilayer, the scan is dominated by the (020) reflection, while the (001) epilayer exhibits the (001), (002), (003), (004), and (005) reflections. The ($\bar{2}01$) epilayer similarly shows the ($\bar{2}01$), ($\bar{4}02$), ($\bar{6}03$), and ($\bar{8}04$) reflections. No additional diffraction peaks associated with secondary $Ga_2O_3$ phases or other orientations were detected, confirming phase-pure homoepitaxial growth on all three substrate orientations. The crystalline quality was further evaluated using ω-rocking curves of the (020), (004), and ($\bar{6}$03) reflections for the (010), (001), and ($\bar{2}$01) epilayers, respectively, as shown in Figs. 3(b)-3(d). The corresponding FWHM values were 63.4, 66.6, and 42.8 arcsec, respectively, demonstrating good crystalline quality across all three orientations. These values are comparable to those reported for high-quality HVPE homoepitaxial β-$Ga_2O_3$ [65, 68, 69, 73].

The influence of growth temperature on the surface morphology of the homoepitaxial β-$Ga_2O_3$ layers was further investigated by FESEM, as shown in Fig. 4. The films were grown at 1050, 1100, and 1150 °C at atmospheric pressure using HCl and $O_2$ flow rates of 400 and 20 sccm, respectively. A pronounced variation in surface morphology is observed with both growth temperature and substrate orientation. For the (010) epilayers [Figs. 4(a)-4(c)], the surface at 1050 °C is characterized by long, parallel features extending along the [001] crystallographic direction. At 1100 °C, the morphology changes considerably, with the formation of wider and more widely separated elongated features. Increasing the temperature to 1150 °C results in a more uniform anisotropic morphology, with broad, continuous parallel features extending along the [001] direction. For the (001) epilayers [Figs. 4(d)-4(f)], the surface at 1050 °C consists of elongated features preferentially aligned along the [010] crystallographic direction. Increasing the temperature to 1100 °C results in a denser distribution of these aligned features, while at 1150 °C, broad, well-defined, and continuous parallel terrace-like features develop along the [010] direction. Similar [010]-aligned anisotropic surface features have been reported for HVPE-grown (001) β-$Ga_2O_3$. Reilly et al. observed pronounced bar-like morphology in thick (001) HVPE epilayers and showed that the as-grown morphology could be substantially modified through optimization of the growth chemistry [65]. A distinctly different surface evolution is observed for the ($\bar{2}01$) orientation [Figs. 4(g)-4(i)]. At 1050 °C, the surface is relatively uniform with sparsely distributed linear features. Increasing the temperature to 1100 °C produces a dense array of closely spaced and highly aligned surface features. At 1150 °C, the morphology develops into broad and well-defined terraces extending over large lateral distances.

To quantitatively evaluate the nanoscale surface morphology, AFM measurements were performed on the (010), (001), and ($\bar{2}01$) epilayers grown at 1150 °C, as shown in Fig. 5. The AFM images reveal a pronounced dependence of the surface morphology on crystallographic orientation. Over the 5 × 5 μm² scan area, the (010) and (001) epilayers exhibit anisotropic elongated surface features with RMS roughnesses of 6.04 and 12.33 nm, respectively. In contrast, the ($\bar{2}01$) epilayer exhibits well-defined step-and-terrace structures with a substantially lower RMS roughness of only 0.49 nm. The corresponding 2 × 2 μm² scans further reveal substantial differences in the local surface morphology. The (010) and (001) epilayers exhibit RMS roughnesses of 7.61 and 4.66 nm, respectively, with the elongated surface features remaining clearly visible. In contrast, the surface within an individual terrace of the ($\bar{2}01$) epilayer is highly smooth, exhibiting an RMS roughness of only 0.17 nm. This sub-nanometer roughness is substantially lower than that obtained for the (010) and (001) epilayers under the same growth conditions, indicating highly smooth local terrace formation on the ($\bar{2}01$) growth plane. The low surface roughness of the ($\bar{2}01$) epilayer is also consistent with its lower rocking-curve FWHM observed in Fig. 3(d).

The influence of HCl flow on the surface morphology of the homoepitaxial layers was also investigated, as shown in Fig. 6. For the (010) epilayers [Figs. 6(a)-6(c)], a strongly anisotropic morphology is maintained throughout the investigated HCl flow range. At 80 sccm HCl, the surface consists of well-defined parallel features extending along the [001] crystallographic direction. With increasing HCl flow, these features become progressively finer and more uniformly distributed while retaining the same preferential alignment. At 400 sccm, closely spaced parallel features extend uniformly across the surface, demonstrating the persistence of the [001]-directed anisotropic morphology over the investigated HCl-flow range. For the (001) epilayers [Figs. 6(d)-6(f)], elongated features aligned along the [010] crystallographic direction are observed throughout the HCl-flow series. The morphology evolves with increasing HCl flow while maintaining the same preferential alignment, with well-defined parallel features remaining evident at 400 sccm. A strong influence of halide precursor supply on (001) $\beta$-$Ga_2O_3$ surface morphology has also been reported, with increasing $Cl_2$ flow from 200 to 400 sccm reducing the surface roughness while the growth rate remained nearly unchanged [65]. The ($\bar{2}01$) epilayers exhibit the most pronounced morphological evolution with HCl flow [Figs. 6(g)-6(i)]. At 80 sccm, the surface is dominated by densely distributed elongated features. Increasing the HCl flow to 300 sccm results in the formation of broader terrace-like regions separated by pronounced linear steps. At 400 sccm, the surface consists of broad terraces separated by widely spaced, well-defined parallel steps. The progressive development of larger terrace regions with increasing HCl flow indicates a strong dependence of the ($\bar{2}01$) surface evolution on HCl supply.

The effect of $O_2$ flow on the surface morphology was further examined at 1150 °C and atmospheric pressure using an HCl flow of 400 sccm, as shown in Fig. 7. As discussed in Fig. 2(a), reducing the $O_2$ flow substantially increases the deposition rate and is accompanied by clear changes in the surface features. At the higher $O_2$ flow, the (010) surface develops finer and more closely spaced [001]-aligned features [Figs. 7(a) and 7(b)], whereas the (001) surface evolves toward broader, well-separated parallel features along [010] [Figs. 7(c) and 7(d)]. The ($\bar{2}01$) orientation retains its characteristic terrace-like morphology at both $O_2$ flows, with broader terraces and more pronounced steps observed at 20 sccm [Figs. 7(e) and 7(f)]. A stronger morphological contrast is observed with growth pressure, as shown in Fig. 8. At 1150 °C, increasing the pressure from 100 to 760 Torr increases the growth rate by more than 2.5 times [Fig. 2(b)] while substantially reorganizing the surface features. The (010) surface changes from relatively uniform regions with sparse elongated features to a dense array of [001]-aligned features. For (001), the fine, closely spaced features at 100 Torr are replaced by broader [010]-aligned features at 760 Torr. The most pronounced transition occurs on (-201), where densely distributed elongated features evolve into broad terraces separated by well-defined parallel steps. These changes likely reflect the pressure-dependent precursor transport and surface incorporation during HVPE growth.

In summary, β-$Ga_2O_3$ homoepitaxy by HVPE was systematically investigated on (010), (001), and (-201) native substrates over a wide range of growth conditions. Increasing the HCl supply and growth temperature substantially enhanced the growth rate, whereas increasing the Ga-source-to-substrate distance reduced the deposition rate. Higher reactor pressure also strongly increased the growth rate, while decreasing the $O_2$ supply resulted in further enhancement. These results demonstrate the strong sensitivity of β-$Ga_2O_3$ HVPE growth kinetics to precursor supply and transport conditions. Phase-pure homoepitaxial layers with good crystalline quality were obtained across all three orientations. Increasing the growth temperature produced pronounced surface evolution, while variations in HCl and $O_2$ supply and reactor pressure further modified the characteristic surface features. The (010) and (001) epilayers consistently exhibited anisotropic features preferentially aligned along [001] and [010], respectively, although their width and spacing varied substantially with the growth conditions. The (-201) surface showed the strongest tendency toward broad, well-defined step-and-terrace structures, particularly with increasing HCl flow and at atmospheric pressure. The (-201) epilayer also exhibited the lowest rocking-curve FWHM together with an exceptionally smooth local surface. These findings demonstrate a wide HVPE growth window for rapid, thick β-$Ga_2O_3$ homoepitaxy and establish the strong dependence of surface evolution on crystallographic orientation and growth conditions.

**Acknowledgement**

This work was supported by the National Science Foundation under Award ECCS-2501623 and Award ECCS-2532898 and Air Force Research Laboratory Regional Network-Mid-Atlantic (AFRL RNMA).

**Data Availability**

The data that support the findings of this study are available from the corresponding author upon reasonable request.

**Conflict of Interest**

The authors have no conflicts to disclose.

**Figure 1**

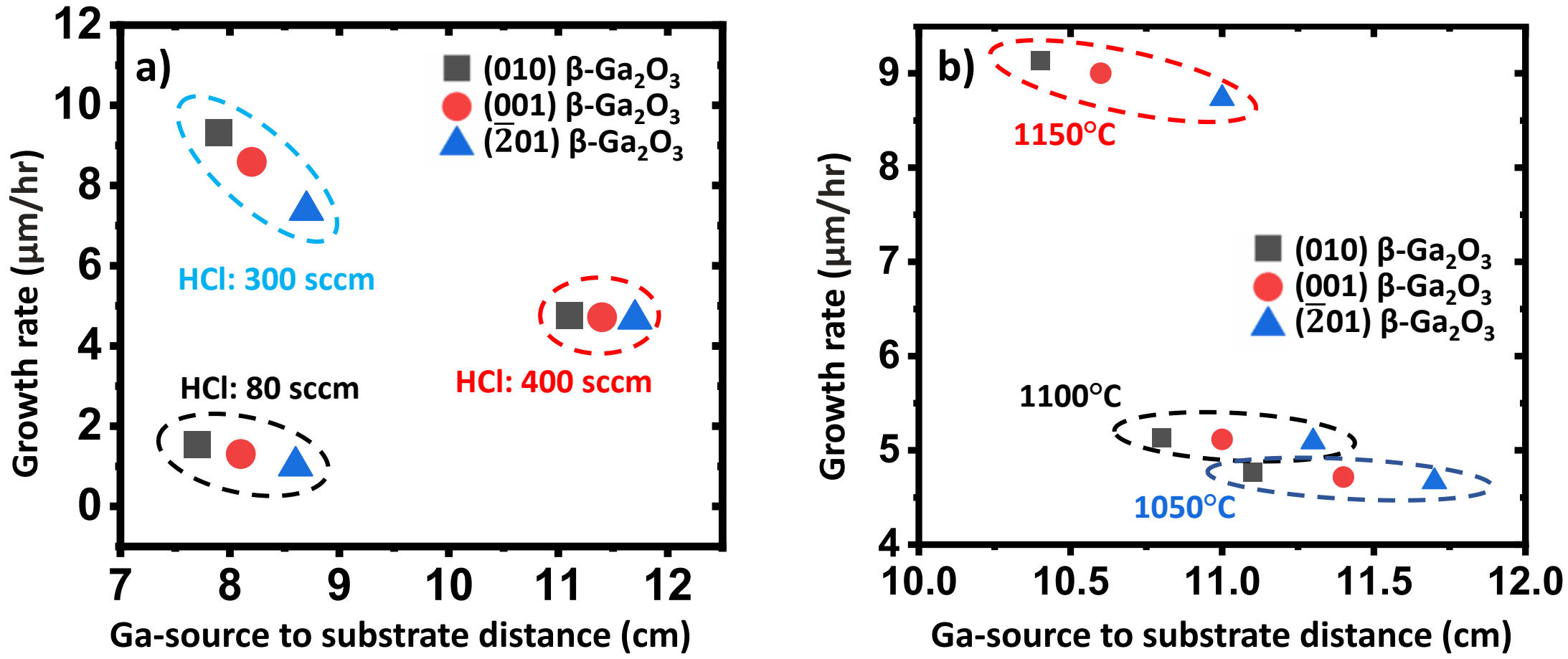

**Figure 1.** Growth rates of β-$Ga_2O_3$ films on (010), (001), and ($\bar{2}$01) substrates as a function of (a) HCl flow and Ga-source-to-substrate distance and (b) growth temperature.

**Figure 2**

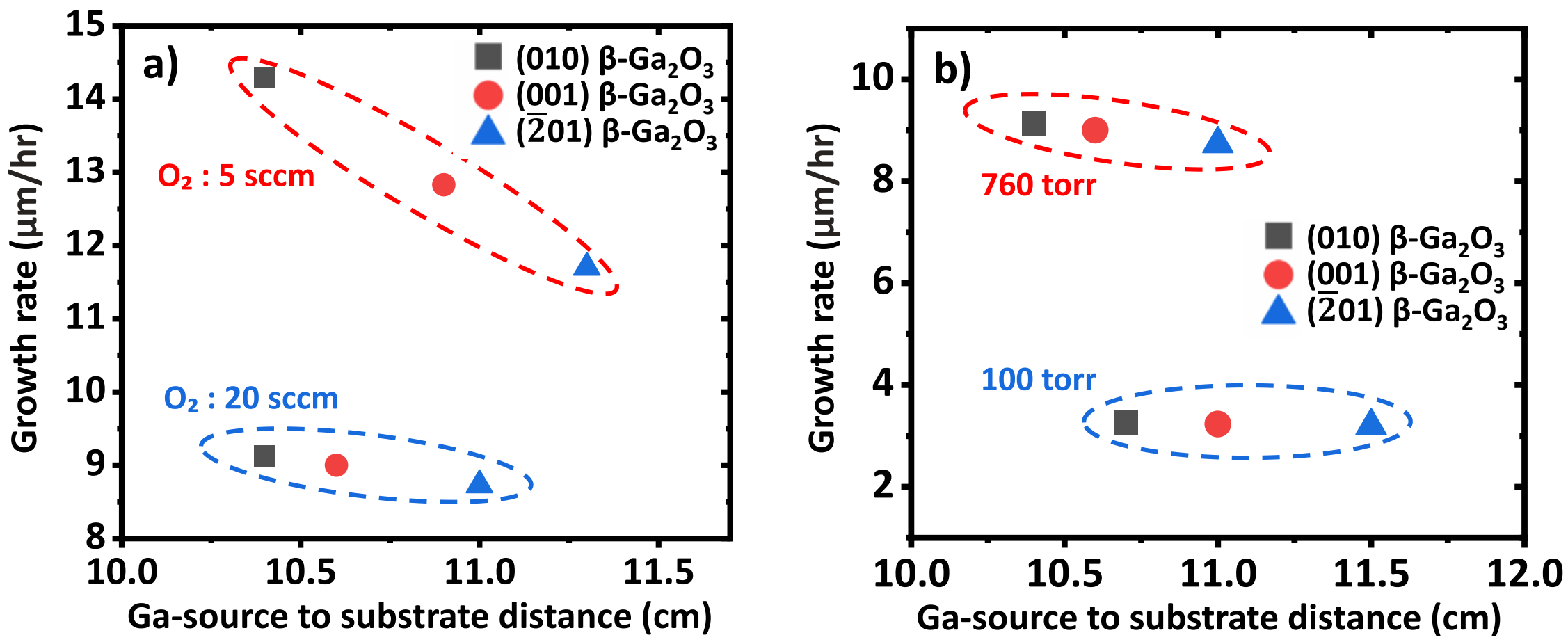

**Figure 2.** Effect of (a) $O_2$ flow and (b) reactor pressure on the growth rates of (010), (001), and ($\bar{2}$01) β-$Ga_2O_3$ epilayers.

Figure 3

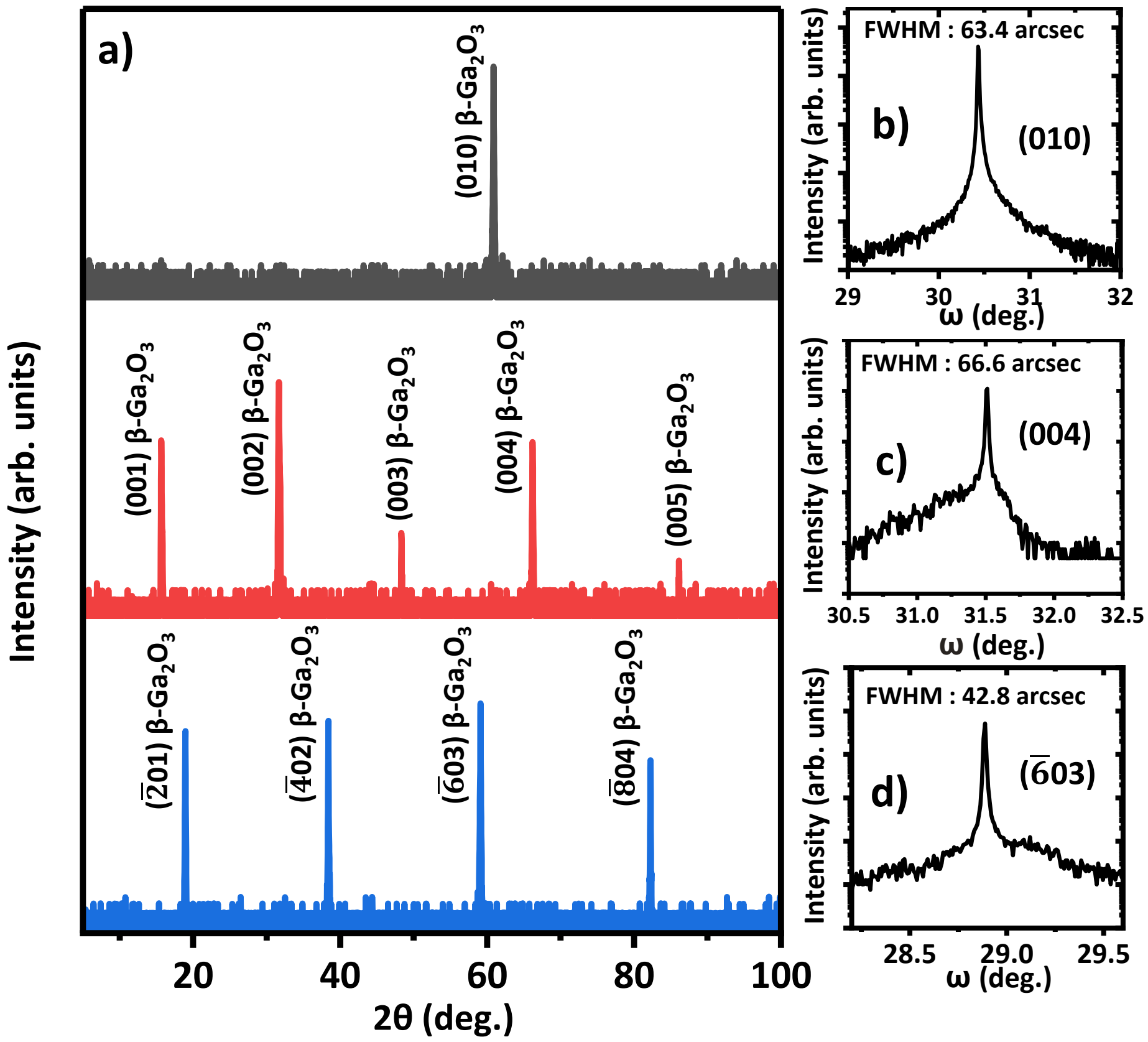

**Figure 3.** (a) ω-2θ XRD scans of the (010), (001), and ($\bar{2}$01) β-$Ga_2O_3$ epilayers and (b-d) corresponding ω-rocking curves, with FWHM values of 63.4, 66.6, and 42.8 arcsec, respectively.

**Figure 4**

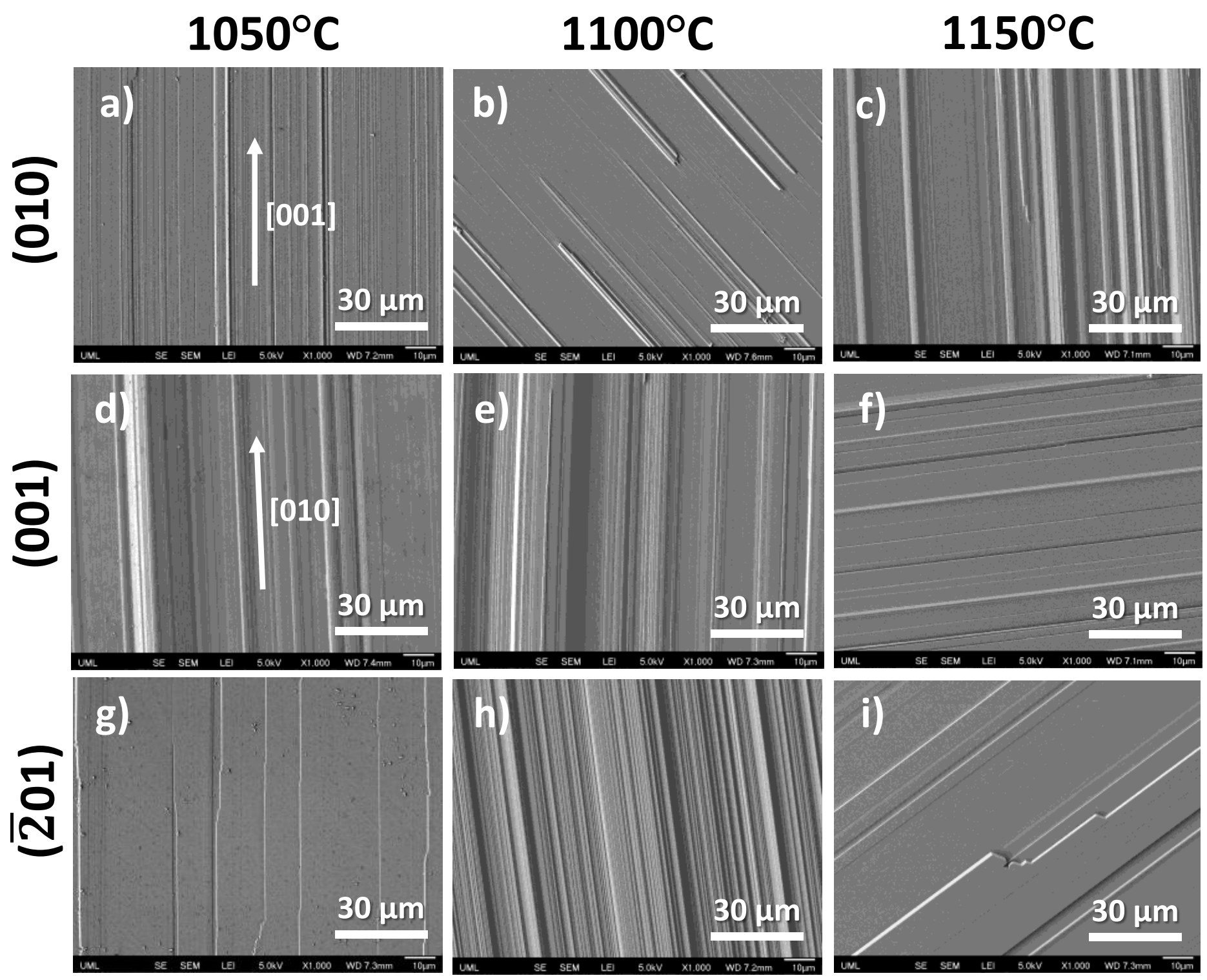

**Figure 4.** Temperature-dependent surface evolution of (a-c) (010), (d-f) (001), and (g-i) ($\bar{2}$01) β-$Ga_2O_3$ epilayers grown at 1050, 1100, and 1150 °C, respectively.

**Figure 5**

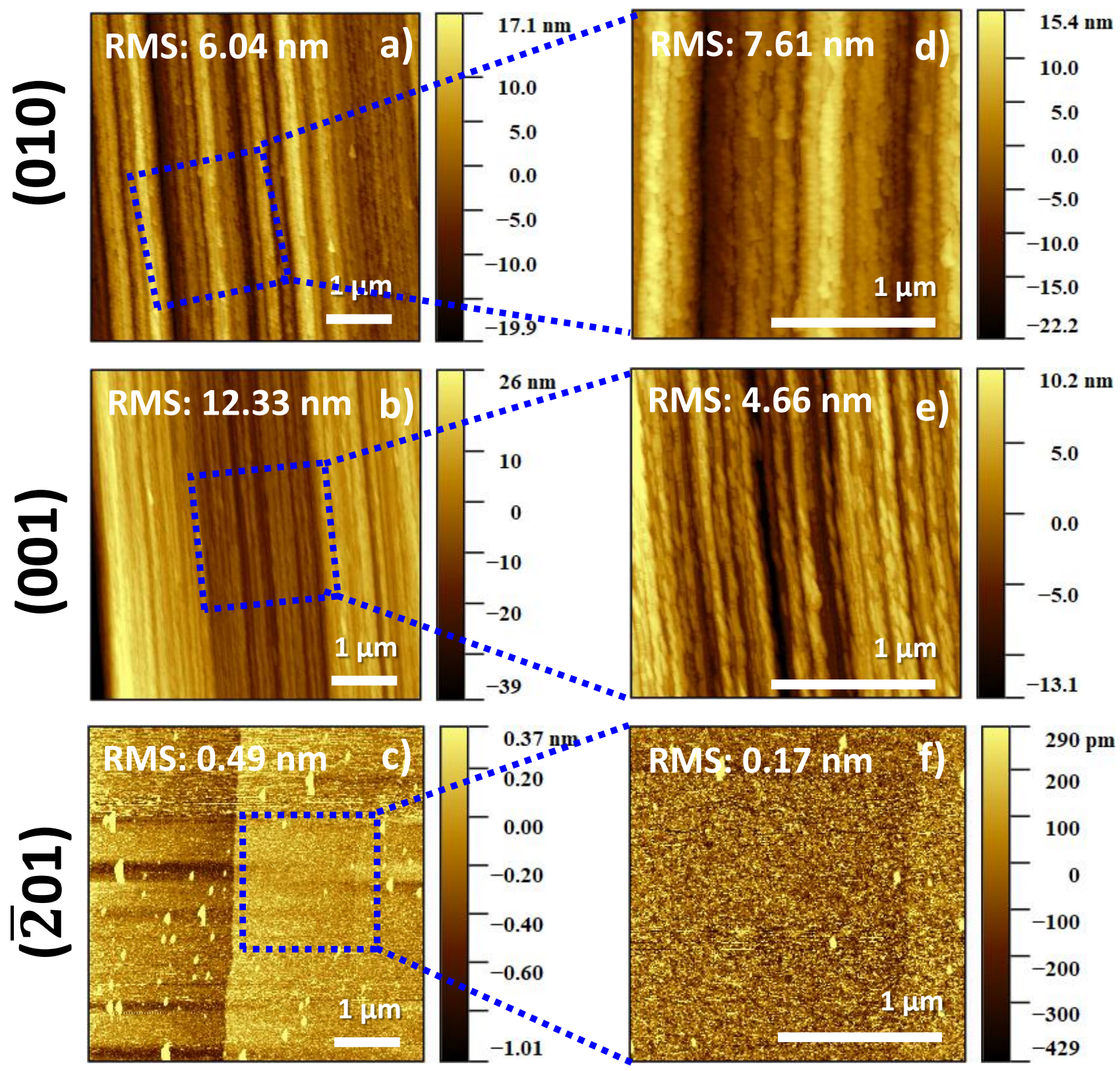

**Figure 5.** AFM surface morphology of (010), (001), and ($\bar{2}01$) β-$Ga_2O_3$ epilayers grown at 1150 °C. (a-c) 5 × 5 μm² scans and (d-f) corresponding 2 × 2 μm² scans, showing the RMS surface roughness for each orientation.

**Figure 6**

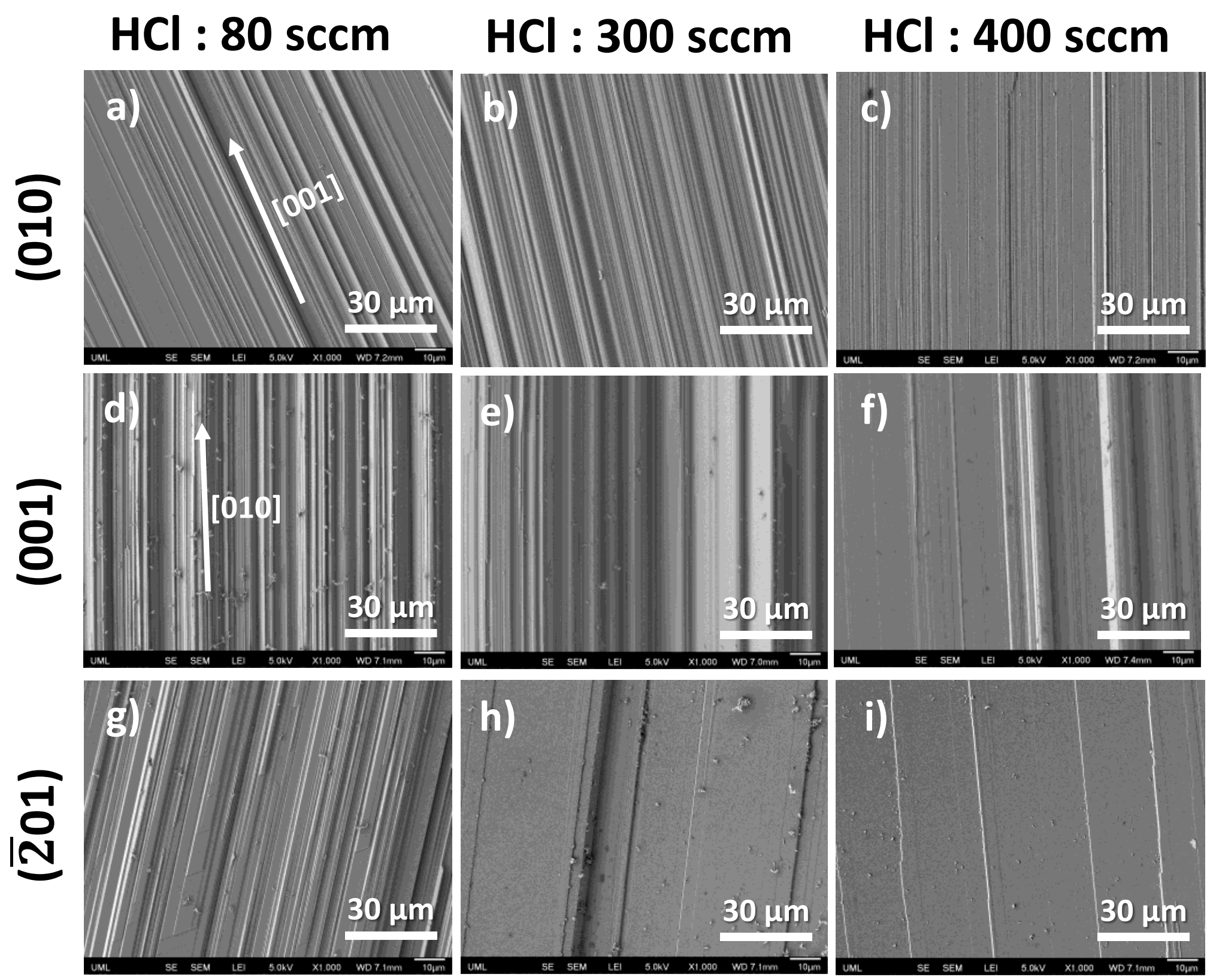

**Figure 6.** HCl-flow-dependent surface evolution of (a-c) (010), (d-f) (001), and (g-i) ($\bar{2}$01) β-$Ga_2O_3$ epilayers at 80, 300, and 400 sccm, respectively.

**Figure 7**

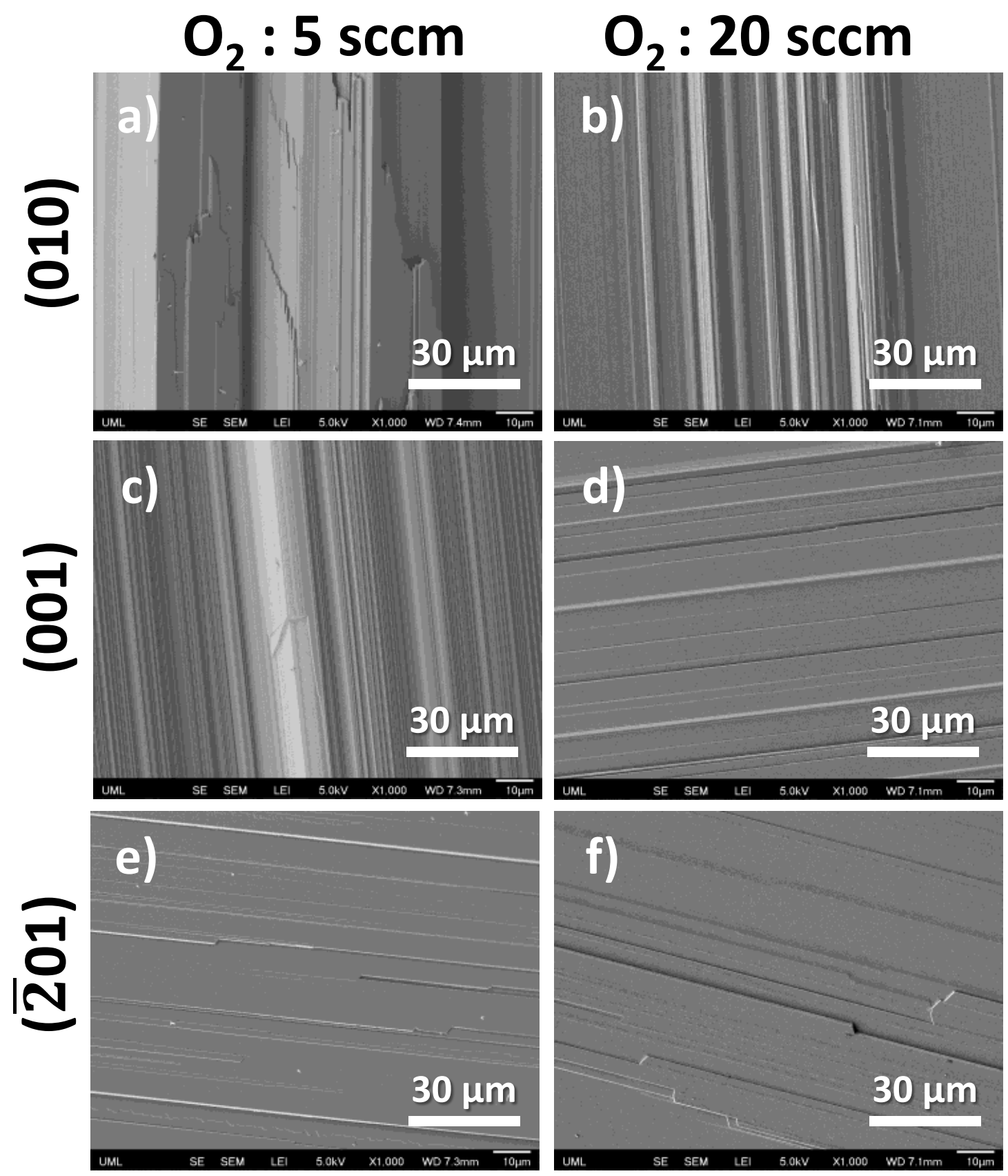

**Figure 7.** Effect of $O_2$ flow on the surface morphology of (a,b) (010), (c,d) (001), and (e,f) ($\bar{2}$01) β-$Ga_2O_3$ epilayers grown at 5 and 20 sccm.

**Figure 8**

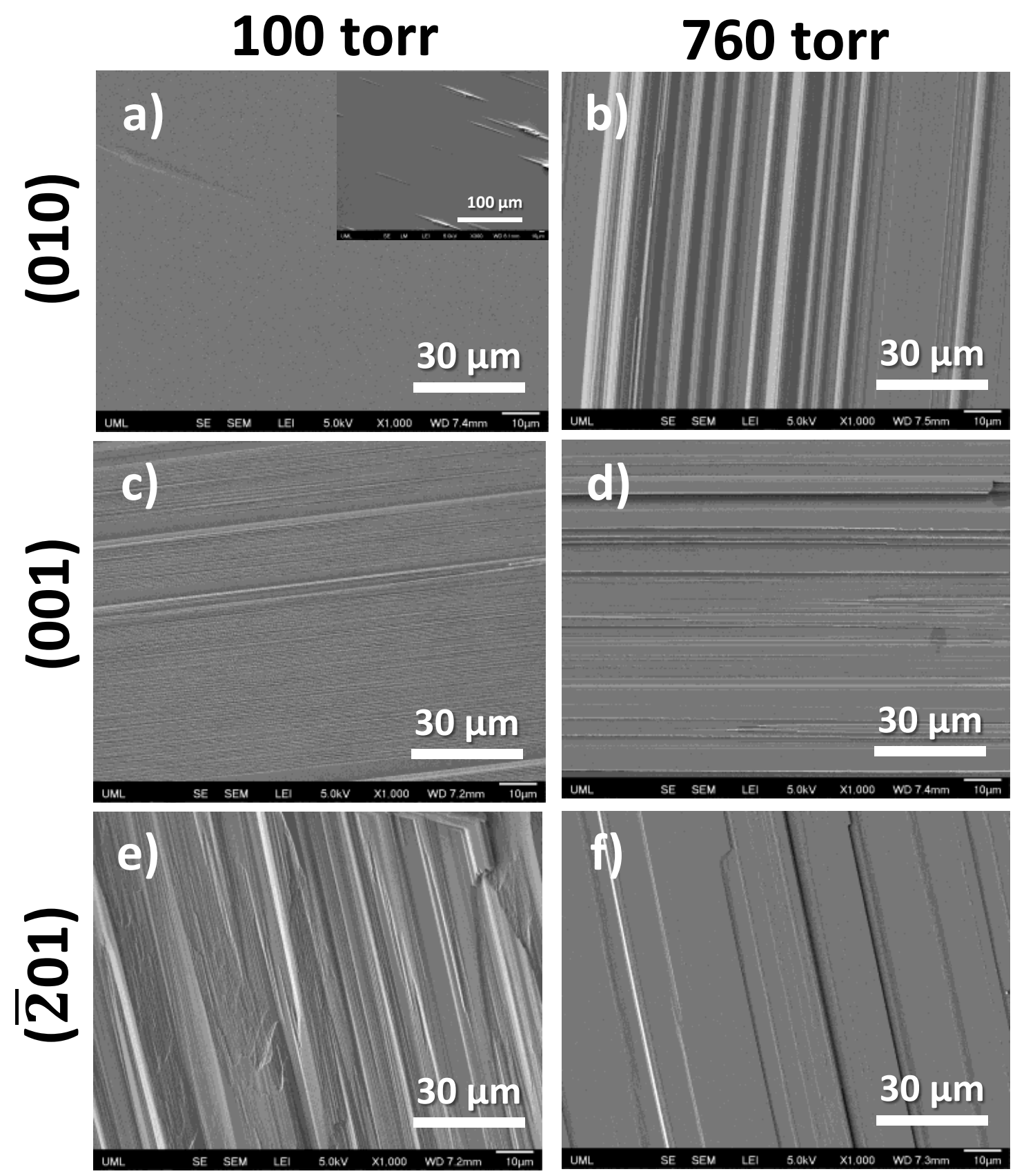

**Figure 8.** Pressure-dependent surface evolution of (a,b) (010), (c,d) (001), and (e,f) ($\bar{2}$01) β-$Ga_2O_3$ epilayers grown at 100 and 760 Torr.